\documentstyle[12pt,epsfig,epsf]{article}

\begin{document}

\title{\Large \bf Differences between Quark and Gluon jets \\as seen at LEP}
\author{Marek Ta\v{s}evsk\'{y} \\
{\small \it CERN, CH-1211 Geneva 23, Switzerland} \\
{\small {\it E-mail:} \sf Marek.Tasevsky@cern.ch}}
\date{}
\maketitle 
\vspace*{-0.8cm}
\begin{center}
{\footnotesize To appear in the {\it Proceedings of the New Trends in 
High-Energy Physics}\\ Yalta, Ukraine, September 22 - 29, 2001.}
\end{center}
\vspace*{0.1cm}
\begin{abstract}\vspace*{0.1cm}
The differences between quark and gluon jets are studied using LEP results
on jet widths, scale dependent multiplicities, ratios of multiplicities, 
slopes and curvatures and fragmentation functions. It is emphasized that the 
observed differences stem primarily from the different quark and gluon colour 
factors.   
\end{abstract}
\vspace*{-12cm}\begin{flushright}
OPAL CR-479\\
31 October 2001
\end{flushright}\vspace*{10cm}
\section{Introduction}
The physics of the differences between quark and gluon jets continuously
attracts an interest of both, theorists and experimentalists. Hadron 
production can be described by parton showers (successive gluon emissions and 
splittings) followed by formation of hadrons which cannot be described 
perturbatively. The gluon emission, being dominant process in the parton 
showers, is proportional to 
the colour factor associated with the coupling of the emitted gluon to the 
emitter. These colour factors are $C_A=3$ when the emitter is a gluon and
$C_F=4/3$ when it is a quark. Consequently, the multiplicity from a gluon 
source is (asymptotically) 9/4 higher than from a quark source.

In QCD calculations, the jet properties are usually defined inclusively, 
by the particles in hemispheres of quark-antiquark ($q\bar{q}$) or gluon-gluon
($gg$) systems in an overall colour singlet rather than by a jet algorithm. 
In contrast to the experimental results which often depend on a jet finder
employed ({\it biased} jets), the inclusive jets do not depend on any jet 
finder ({\it unbiased} jets).
\vspace*{-0.2cm}
\section{Results}
\vspace*{-0.1cm}
\subsection{Jet Widths}
\begin{figure}[h]\centering
\epsfig{file=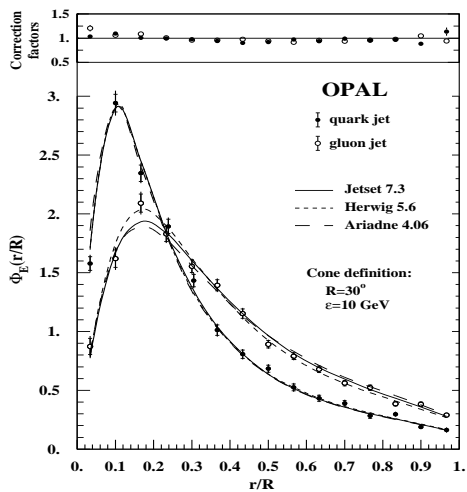,bbllx=87pt,bblly=105pt,bburx=524pt,bbury=680pt,%
width=6cm,height=6.5cm,clip=}
\vspace*{-0.2cm}
\caption{\small The differential energy profile of gluon and quark jets defined
using a cone jet algorithm \cite{jetwidth}.}
\label{jet_width}
\end{figure}

As a consequence of the greater radiation of soft gluons in a gluon jet 
compared to a quark jet, gluon jets are predicted to be broader. An 
experimental confirmation of this effect is shown in Fig.\ref{jet_width} where
the fraction of a jet's visible energy close to the jet axis is larger for 
quark jets than for gluon jets. All the QCD-based models describe the data 
very well.

\vspace*{-0.4cm}
\subsection{Multiplicity Distributions and Ratios}
\begin{figure}[h]\centering
\epsfig{file=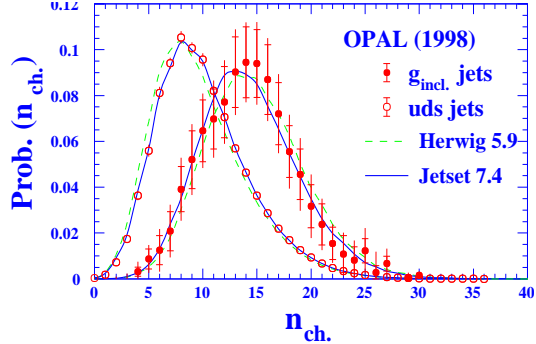,bbllx=0pt,bblly=0pt,bburx=450pt,bbury=280pt,%
width=7cm,height=4.5cm,clip=}
\vspace*{-0.2cm}
\caption{\small Charged particle multiplicity of unbiased gluon and uds 
flavoured jets \cite{gincl1}.}
\label{multipl}
\end{figure}

The predicted larger soft gluon emission in gluon jets compared to quark jets 
has been confirmed by an observed difference between the hadron multiplicity
in quark and gluon jets where the latter are found to be higher, as can be seen
for example in Fig.\ref{multipl} \cite{gincl1}. Only unbiased jets (here 
$g_{incl.}$) defined by particles found in the event hemispheres were used. 
The hemispheres are defined by the plane perpendicular to the principal event 
axis.
There is a large theoretical interest in the ratio of the
mean multiplicity of gluon and quark jets, $r=\langle N_g\rangle/\langle N_q\rangle$. This is predicted to be equal to the ratio $C_A/C_F=2.25$ if the 
asymptotic condition $E_{particle}\ll E_{jet}$ is fulfilled. In real 
experimental conditions ($E_{jet}$ finite), the satisfaction of this condition
is approached by taking only soft particles into account \cite{KhLuOc}. 
In \cite{gincl2} soft particles in unbiased gluon and quark jets 
($E_{jet}\!\sim\!40$ GeV) were defined by momenta $p<$2.0 GeV. 
In order to reduce the hadronization effects, transverse momenta 
of particles relative to the jet axes were required to be higher than 0.8 
GeV, yielding $r=2.32\pm0.18$ which agrees with the asymptotic value. 
The corresponding HERWIG results for $E_{c.m.}=91$ GeV were found to be in a 
good agreement with the measurement. Moreover, for asymptotic $E_{c.m.}=10$ 
TeV, HERWIG yielded $r=2.25$, while JETSET set to have $C_A\!=\!C_F\!=\!4/3$ 
gave $r=1.00$.

Exploiting all the particles from finite energy jets leads to a reduced 
value of $r$ compared to the asymptotic one. The measured value 1.51$\pm$0.04
from \cite{gincl1} is in excellent agreement with QCD calculations of this 
quantity \cite{LuOc,EdGu}. 
\vspace*{-0.2cm}
\subsection{Scale Dependent Multiplicities and Ratios}\vspace*{-0.1cm}
Adopting a recently proposed method for obtaining the scale dependent unbiased
gluon jet multiplicity, $N_g(Q)$ \cite{EdGu,EdGuKh}, the ratios of
multiplicities, $r$, of slopes, $r^{(1)}$ and of curvatures, $r^{(2)}$, 
defined as\vspace*{-0.3cm}\\
\begin{equation}
r^{(1)} \equiv \frac{{\mathrm dN}_g/{\mathrm d}y}
                     {{\mathrm dN}_q/{\mathrm d}y}, \hspace*{0.3cm}
r^{(2)} \equiv \frac{{\mathrm d^2N}_g/{\mathrm d^2}y}
                     {{\mathrm d^2N}_q/{\mathrm d^2}y},\hspace*{0.5cm} 
y = \ln (Q/\Lambda), \hspace*{0.2cm} Q=E_{jet}
\label{rdef}
\end{equation}
were recently measured \cite{OPind,DEind}
and compared to recent QCD calculations \cite{LuOc,EdGu,Cap}. 
The method is based on a NLO expression for $N_{gg}$:\vspace*{-0.2cm}
\begin{equation}
N_{gg}^{ch}(k_{\bot,Lu}) = 2[N_{q\bar{q}g}^{ch} 
-N_{q\bar{q}}^{ch}(L,k_{\bot,Lu})]
\label{NgLu}
\vspace*{-0.2cm}
\end{equation}
\begin{equation}
N_{gg}^{ch}(k_{\bot,Le}) = 2[N_{q\bar{q}g}^{ch} 
-N_{q\bar{q}}^{ch}(L_{q\bar{q}},k_{\bot,Lu})]
\label{NgLe}
\end{equation}
where $N_{gg}$ is the inclusive multiplicity in 2-jet $gg$ system and
$N_{q\bar{q}}$ is the exclusive multiplicity in 2-jet $q\bar{q}$ events with
no gluon radiation harder than $k_{\bot,Lu}$.
$N_{q\bar{q}g}$ is the multiplicity of e$^+$e$^-$ 3-jet events. The two
expressions for $N_{gg}$ reflect the ambiguity in the definition of the gluon
jet $p_{\bot}$ with respect to the $q\bar{q}$ system when the gluon radiation
is hard. The scales $k_{\bot,Lu}$ and $k_{\bot,Le}$ are proportional to\vspace*{-0.2cm}\\
\begin{equation}\vspace*{-0.2cm}
p_{\bot,Lu} = \sqrt{\frac{s_{qg}s_{\bar{q}g}}{s}},\hspace*{0.5cm}
p_{\bot,Le} = \sqrt{\frac{s_{qg}s_{\bar{q}g}}{s_{q\bar{q}}}}
\label{pt}
\end{equation} 
with $s=E_{c.m.}^2$, $s_{q\bar{q}}=p_qp_{\bar{q}}$, $s_{{q}g}=p_qp_g$,
$s_{{\bar{q}}g}=p_{\bar{q}}p_g$ and $p_q, p_{\bar{q}}$ and $p_g$ the 4-momenta
of the $q,\bar{q}$ and $g$. $L$ specifies the e$^+$e$^-$ c.m. energy 
($E_{c.m.}$) and $L_{q\bar{q}}$ the energy of the $q\bar{q}$ system in the 
$q\bar{q}$ rest frame. Note that the gluon jet terms depend on a single scale
which corresponds to the unbiased jets, whereas the quark jet terms depend on
two scales accounting for the bias in quark jet multiplicity due to the jet
finder criteria used to select the $q\bar{q}g$ events.  

\begin{figure}[h]\centering
\epsfig{file=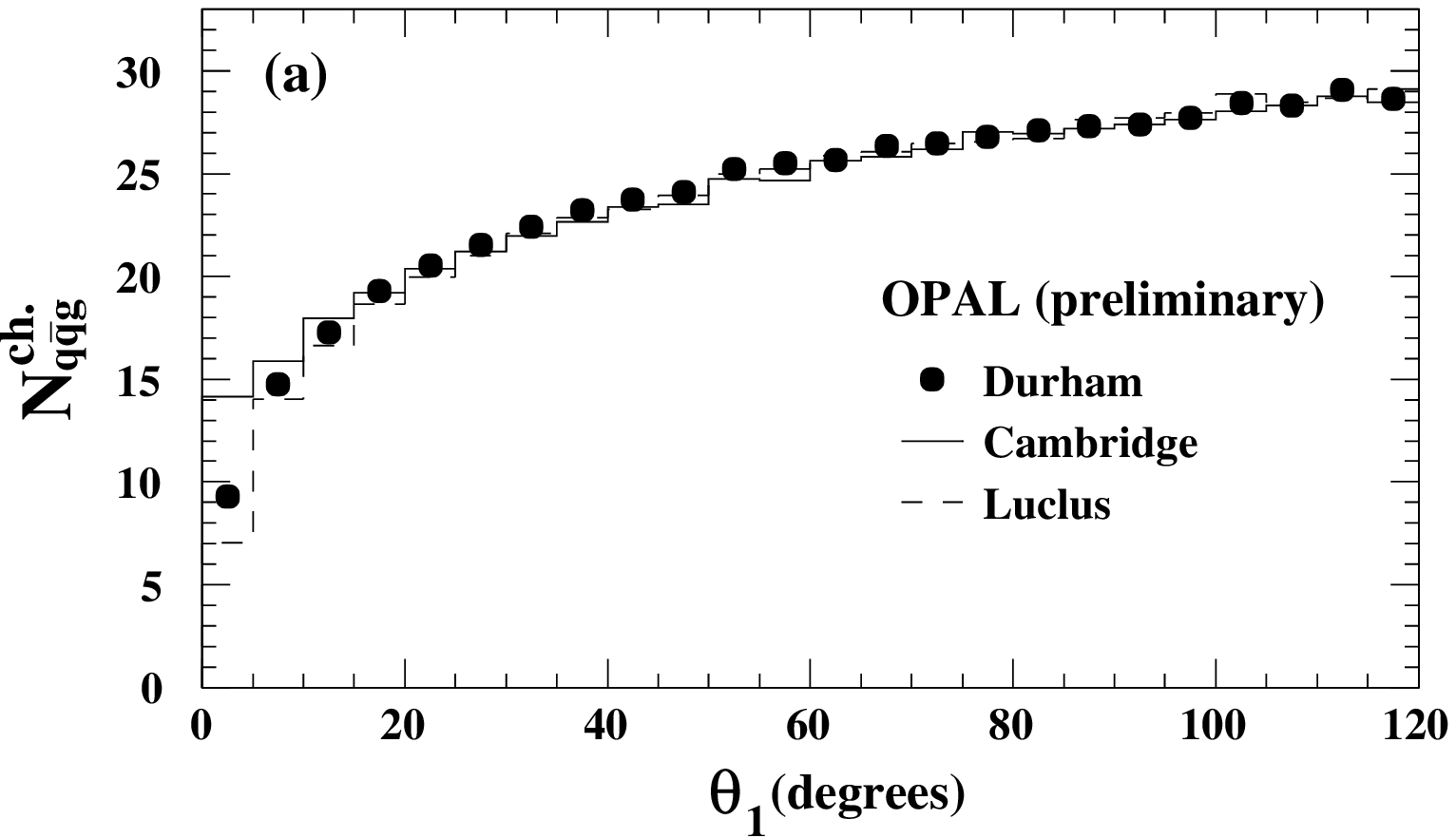,bbllx=0pt,bblly=0pt,bburx=440pt,bbury=253pt,%
width=6.8cm,height=4cm,clip=}
\epsfig{file=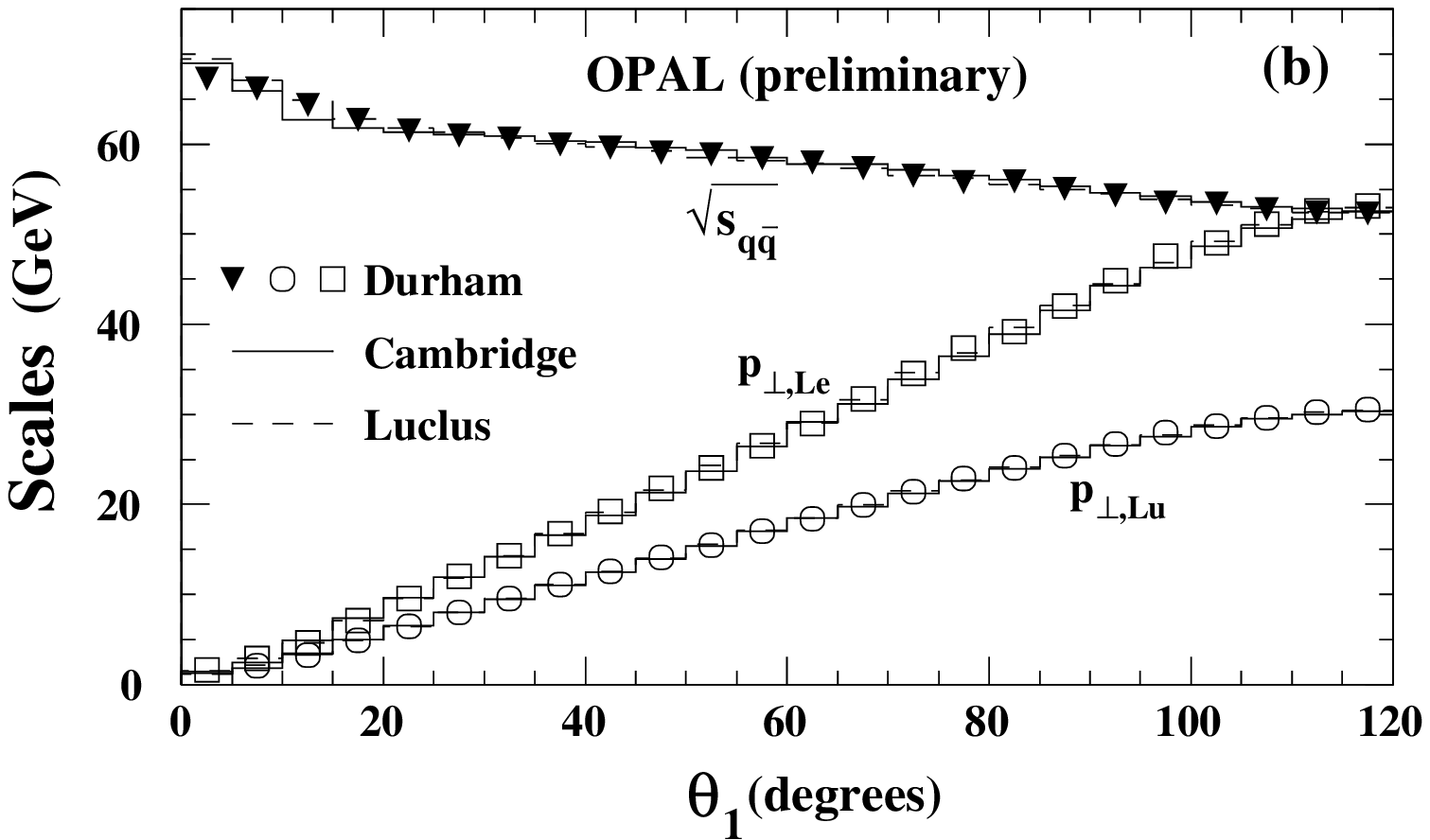,bbllx=0pt,bblly=0pt,bburx=440pt,%
bbury=255pt,width=6.8cm,height=4cm,clip=}
\vspace*{-0.1cm}
\caption{\small (a) The mean charged particle multiplicity of 3-jet uds 
flavour Y events from $Z^0$ decays, selected using the Durham, Cambridge and 
Luclus jet finders as a function of the opening angle $\theta_1$. (b) The 
corresponding scales defined in Eq.\ref{pt}. }
\label{Yevents}
\end{figure}

In order to obtain $N_{gg}$, two event samples with jets found by Durham,
Cambridge and Luclus jet finders were used. In the first sample,
3-jet light quark (uds) events\footnote{Theoretical expressions are based on 
massless quarks} from $Z^0 \rightarrow q\bar{q}$ decays were kept. 
After energy ordering, the jet 3, having the lowest energy, is
taken to be the gluon jet. This fact together with the condition 
$\theta_2\!\approx\!\theta_3$ (the angles between the jet 1 and the other two 
are roughly the same, so called ``Y events'') leads to a low sensitivity to
gluon jet mis-identification. For Y events, the quantities depend only on 
$E_{c.m.}$ and one inter-jet angle, which was conveniently chosen to be 
$\theta_1$. The measurement of $N^{ch}_{q\bar{q}g}, L_{q\bar{q}}, 
k_{\bot,Lu}$ and $k_{\bot,Le}$ is shown in Fig.\ref{Yevents}.\\  
In the second sample, $N^{ch}_{q\bar{q}}(L,k_{\bot,Lu})$ from 
Eq.\ref{NgLu} was directly measured as a mean multiplicity of 2-jet uds flavour
events from $Z^0$ decays ($L=\ln(M_Z^2/\Lambda^2$) fixed). Note that 
$N^{ch}_{q\bar{q}}(L_{q\bar{q}},k_{\bot,Lu})$ cannot be directly measured 
since $L_{q\bar{q}}$, unlike $L$, is variable, so a direct measurement 
relevant for this analysis would require c.m. energies below the $Z^0$. 
Instead, the biased $N^{ch}_{q\bar{q}}$ is determined from measurements of the
unbiased $N^{ch}_{q\bar{q}}$, using a NLO expression from \cite{EdGu}.\\
\vspace*{0.2cm}
\begin{figure}[h]\centering
\epsfig{file=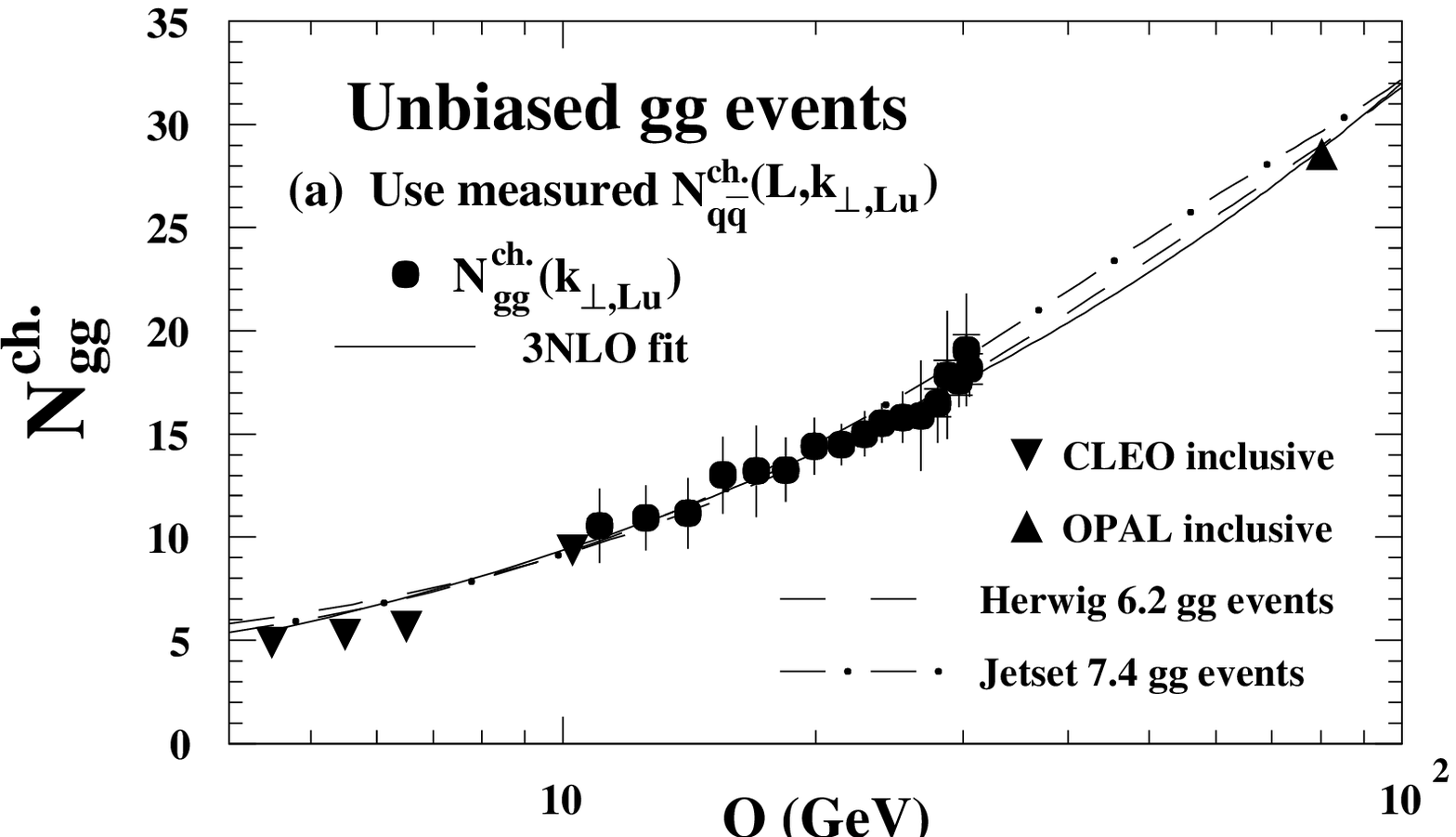,bbllx=0pt,bblly=-4pt,bburx=455pt,bbury=260pt,%
width=6.8cm,height=3.8cm,clip=}
\epsfig{file=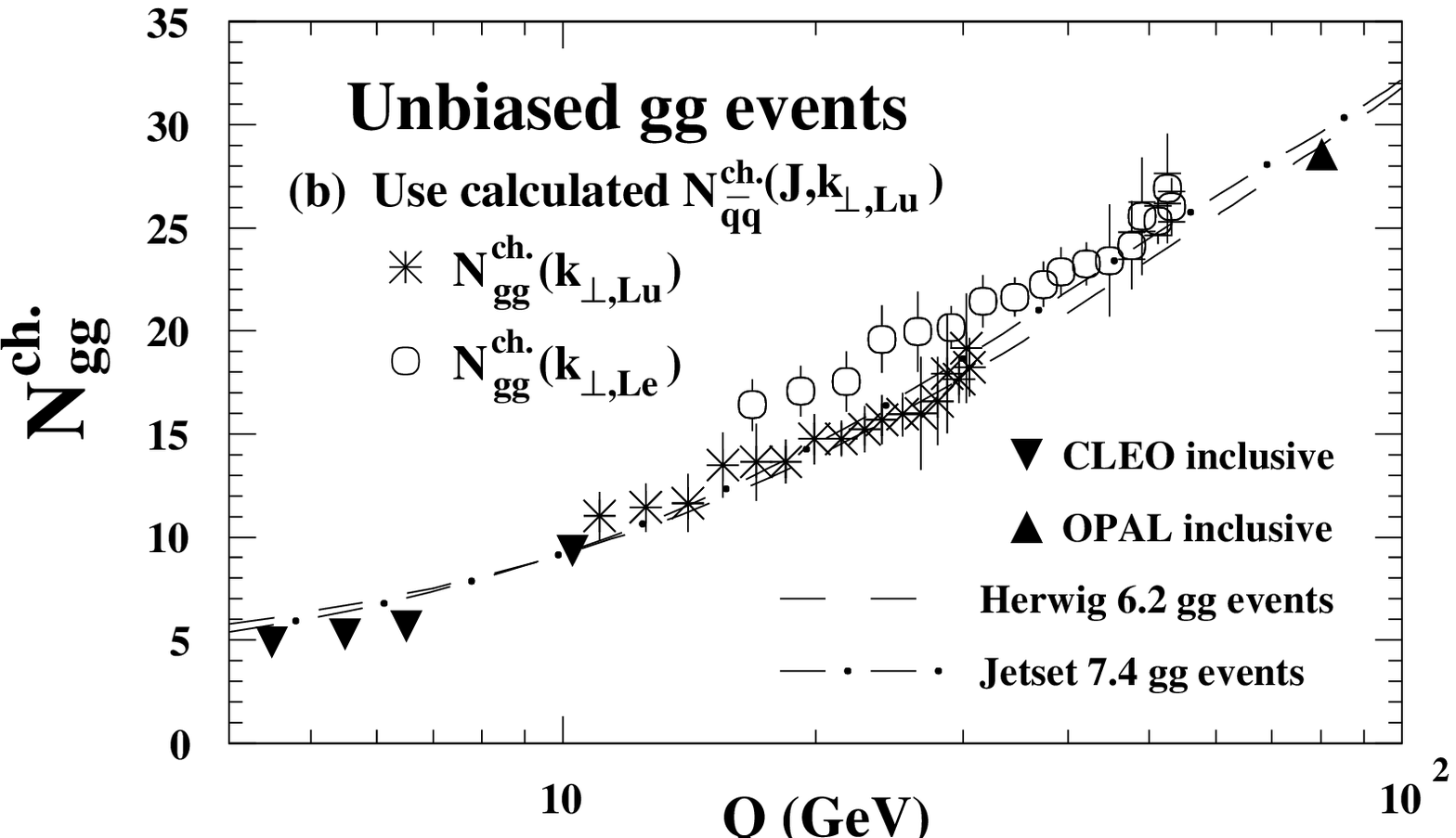,bbllx=0pt,bblly=-4pt,bburx=455pt,bbury=260pt,%
width=6.8cm,height=3.8cm,clip=}
\vspace*{-0.2cm}
\caption{\small The mean multiplicity of unbiased $gg$ events as a function 
of scale. (a) Results from Eq.\ref{NgLu} using the measured 
$N^{ch}_{q\bar{q}}(L,k_{\bot,Lu})$. (b) The corresponding results from 
eqs.\ref{NgLu} and \ref{NgLe} using the NLO expression for the biased 
$N_{q\bar{q}}$ \cite{EdGu}. The triangles show CLEO \cite{CLEO} and OPAL 
\cite{gincl2} measurements of inclusive unbiased multiplicity.}
\label{Ngg}
\vspace*{-0.4cm}
\begin{picture}(0,0)(0,0)
\put(-20,175){\bf Preliminary}
\end{picture}
\end{figure}
Fig.\ref{Ngg} shows that the results for $N_{gg}$ using the directly measured 
biased 2-quark jet multiplicity at Lund definition of $k_{\bot}$ 
are found consistent with the direct CLEO \cite{CLEO} and OPAL \cite{gincl2} 
measurements and with 
MC predictions (Fig.\ref{Ngg}(a)) as well as with the result using the 
calculated $N^{ch}_{q\bar{q}}(L,k_{\bot,Lu})$ (Fig.\ref{Ngg}(b)). 
On the other hand, using the Leningrad definition of $k_{\bot}$, the results 
(Fig.\ref{Ngg}(b)) are inconsistent with MC predictions (found to have been  
accurate for $N_{q\bar{q}}$ and $N_{gg}$ in many other studies) and also with 
direct CLEO (and possibly OPAL) measurements. 
These observations show a clear preference of the results based on 
Eq.\ref{NgLu} using $k_{\bot,Lu}$ over those based on $k_{\bot,Le}$.
\begin{figure}[h]\centering
\epsfig{file=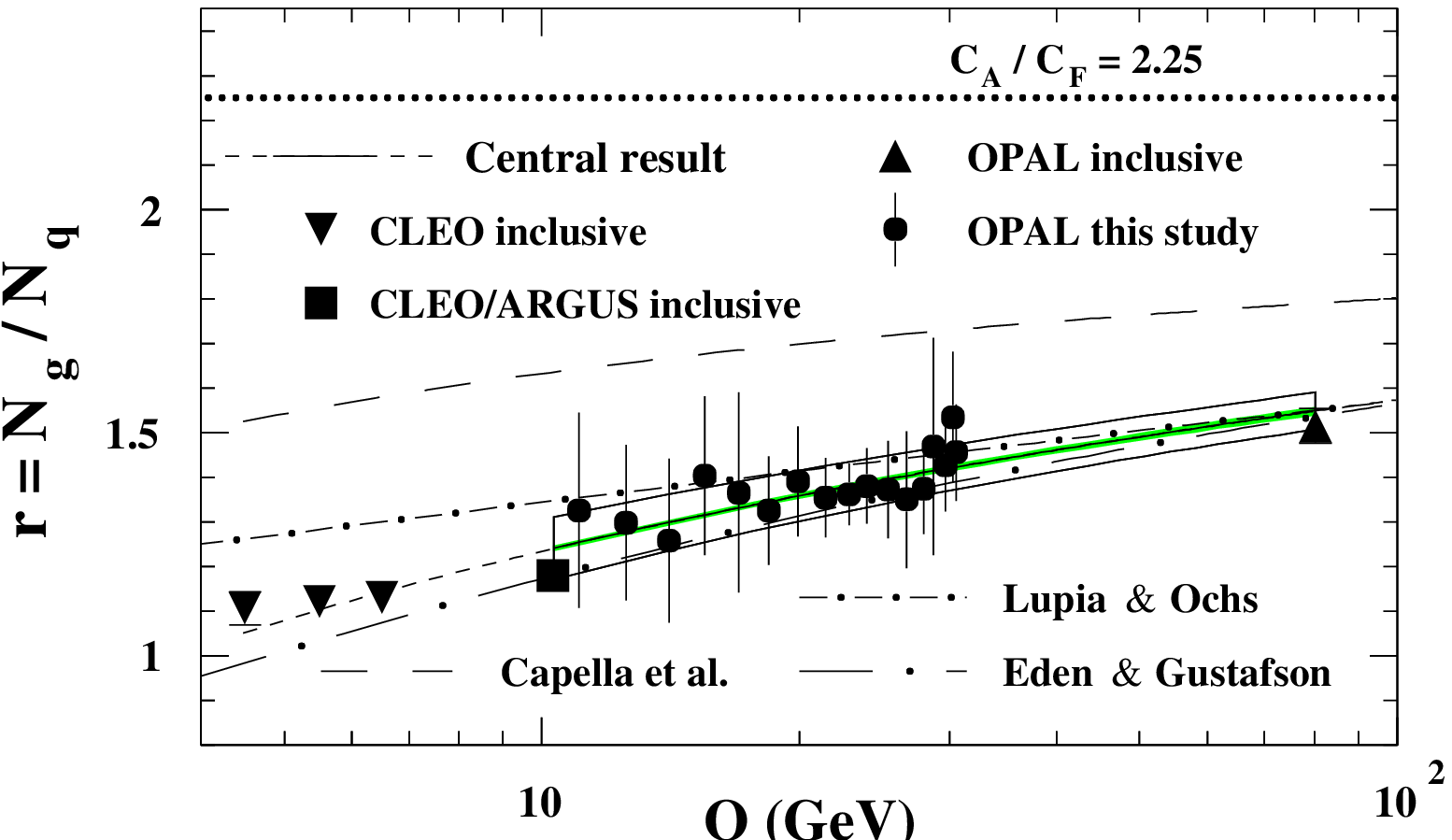,bbllx=0pt,bblly=-4pt,bburx=447pt,bbury=258pt,%
width=5.8cm,height=3.5cm,clip=}
\epsfig{file=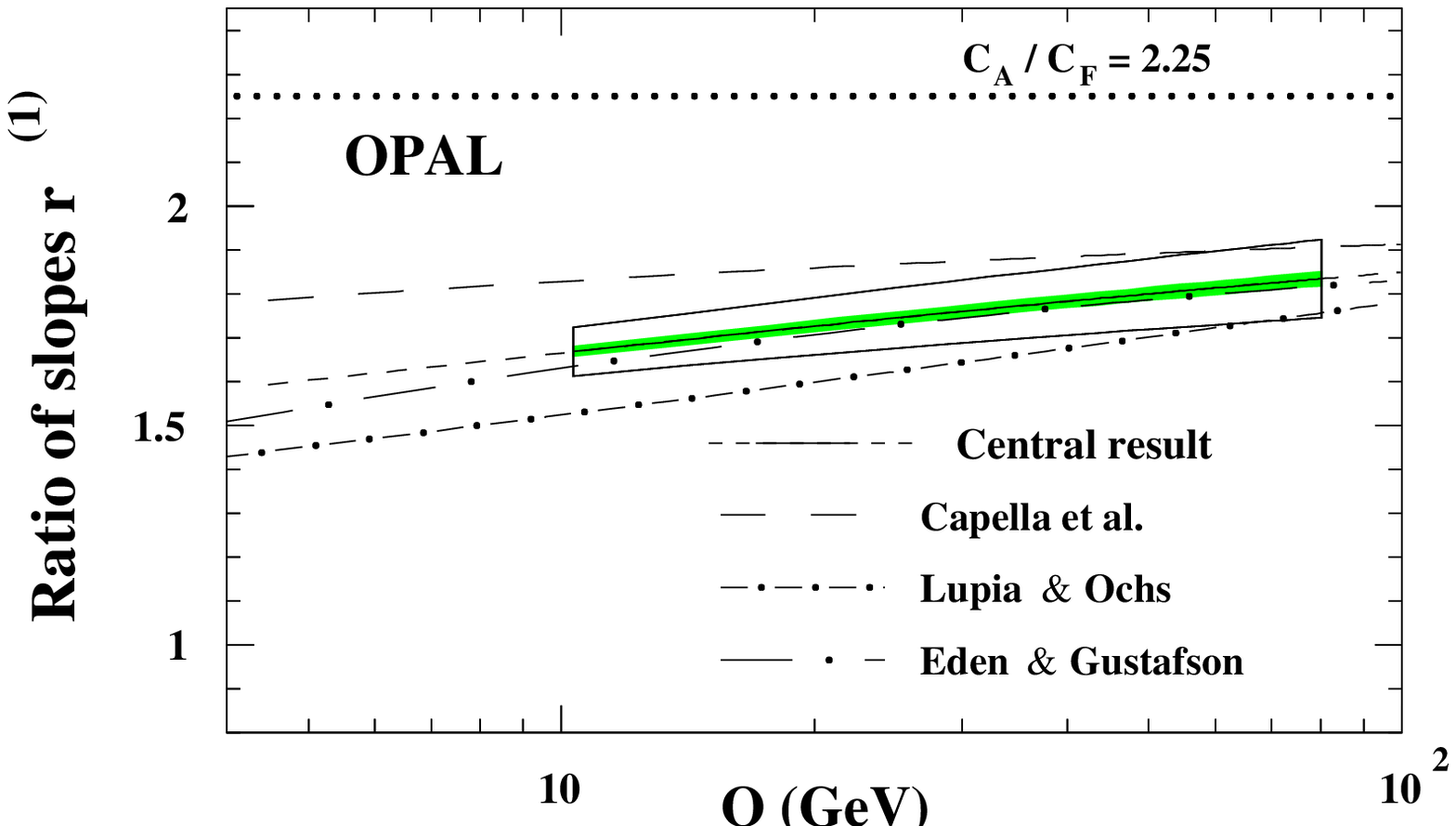,bbllx=0pt,bblly=-4pt,bburx=447pt,bbury=258pt,%
width=5.8cm,height=3.5cm,clip=}
\epsfig{file=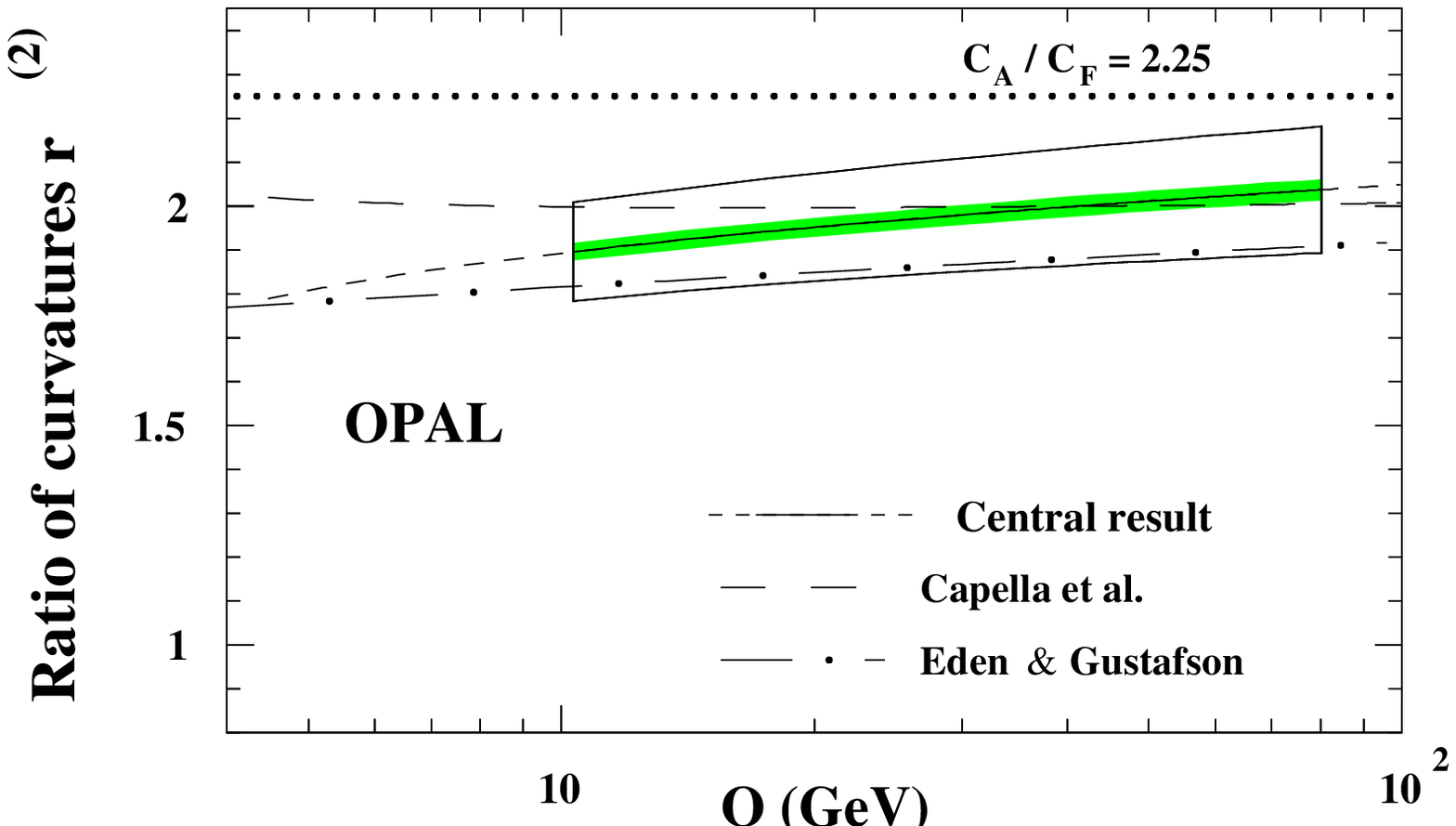,bbllx=0pt,bblly=-4pt,bburx=447pt,bbury=258pt,%
width=5.8cm,height=3.5cm,clip=}
\vspace*{-0.1cm}
\caption{\small The ratios of the mean multiplicity $r$, of slopes $r^{(1)}$ 
and of curvatures $r^{(2)}$ between unbiased gluon and uds quark jets as a
function of scale.}
\label{rr1r2}
\vspace*{-0.4cm}
\end{figure}
In Fig.\ref{rr1r2} the data are compared to various QCD calculations 
\cite{LuOc,EdGu,Cap}. For the predictions of \cite{Cap} we observe that at 30 
GeV, $r$ and $r^{(1)}$ exceed the data by about 20 and 6\%, while $r^{(2)}$ 
agrees with the data. This suggests that higher order corrections are smaller 
for $r^{(2)}$ than for $r^{(1)}$ and for $r^{(1)}$ than for $r$. The data 
also confirm the prediction 
$r\!<\!r^{(1)}\!<\!r^{(2)}\!<\!C_A/C_F\!=\!2.25$. For the predictions of 
\cite{LuOc} we observe a better agreement than \cite{Cap} for $r$ and a 
similar agreement for $r^{(1)}$. The predictions of \cite{EdGu} are in good 
overall agreement with the data, however, it should be noted that these 
predictions are not entirely independent of the data.
\vspace*{-0.35cm}
\subsection{Fragmentation Functions}
The differences between quark and gluon jets manifest themselves also in
the fragmentation functions, defined as\vspace*{-0.1cm}\\
\begin{equation}
D(x_E,Q) \equiv \frac{1}{N_{jet}(Q)}\frac{{\mathrm d}N_{part}(x_e,Q)}
{{\mathrm d}x_E}, \hspace*{0.5cm} x_E=\frac{E_{part}}{E_{jet}}
\label{ff}
\end{equation}
\begin{picture}(0,0)(0,0)
\put(180,355){\bf Preliminary}
\end{picture} 
In \cite{Delff} the quark and gluon fragmentation functions have been measured
in udsc flavour general as well as in Y 3-jet events using Durham and 
Cambridge jet finders. The scale in
these cases is not unambiguously defined but it should depend on  
$E_{jet}$ and the event topology. Studies of hadron production in events with
a general topology have shown that the characteristics of the parton cascade
depend mainly on the hardness of the process producing jets 
\cite{Dok}\vspace*{-0.1cm}\\
\begin{equation}
\kappa_H=E_{jet}\sin\theta/2, \hspace*{0.5cm} \theta={\mathrm angle~to~the~closest~jet}
\label{scale}
\end{equation}
and accordingly, the scale in this analysis was put $Q=\kappa_H$. 
\begin{figure}\centering
\epsfig{file=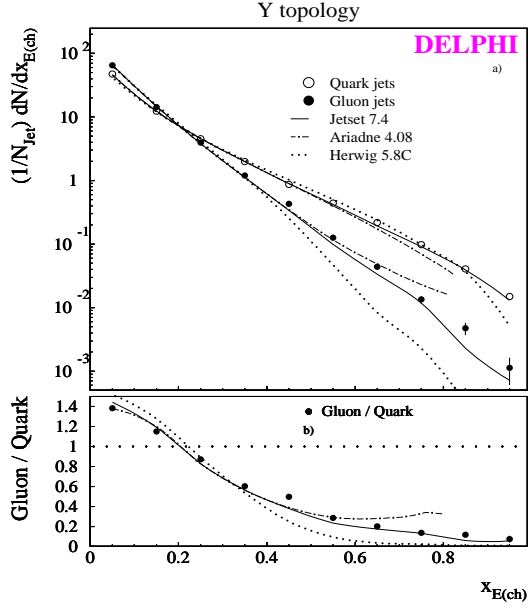,width=7cm,height=8cm,%
bbllx=45pt,bblly=115pt,bburx=545pt,bbury=718pt,clip=}
\caption{\small Quark and gluon jet fragmentation functions of Y events,
$\theta_2,\theta_3\!\in\![150^\circ\!\pm15\!^\circ]$, compared to the 
predictions of various fragmentation models (Durham alg.).}
\label{ffincl}
\end{figure}
\begin{figure}\vspace*{-0.3cm}
\epsfig{file=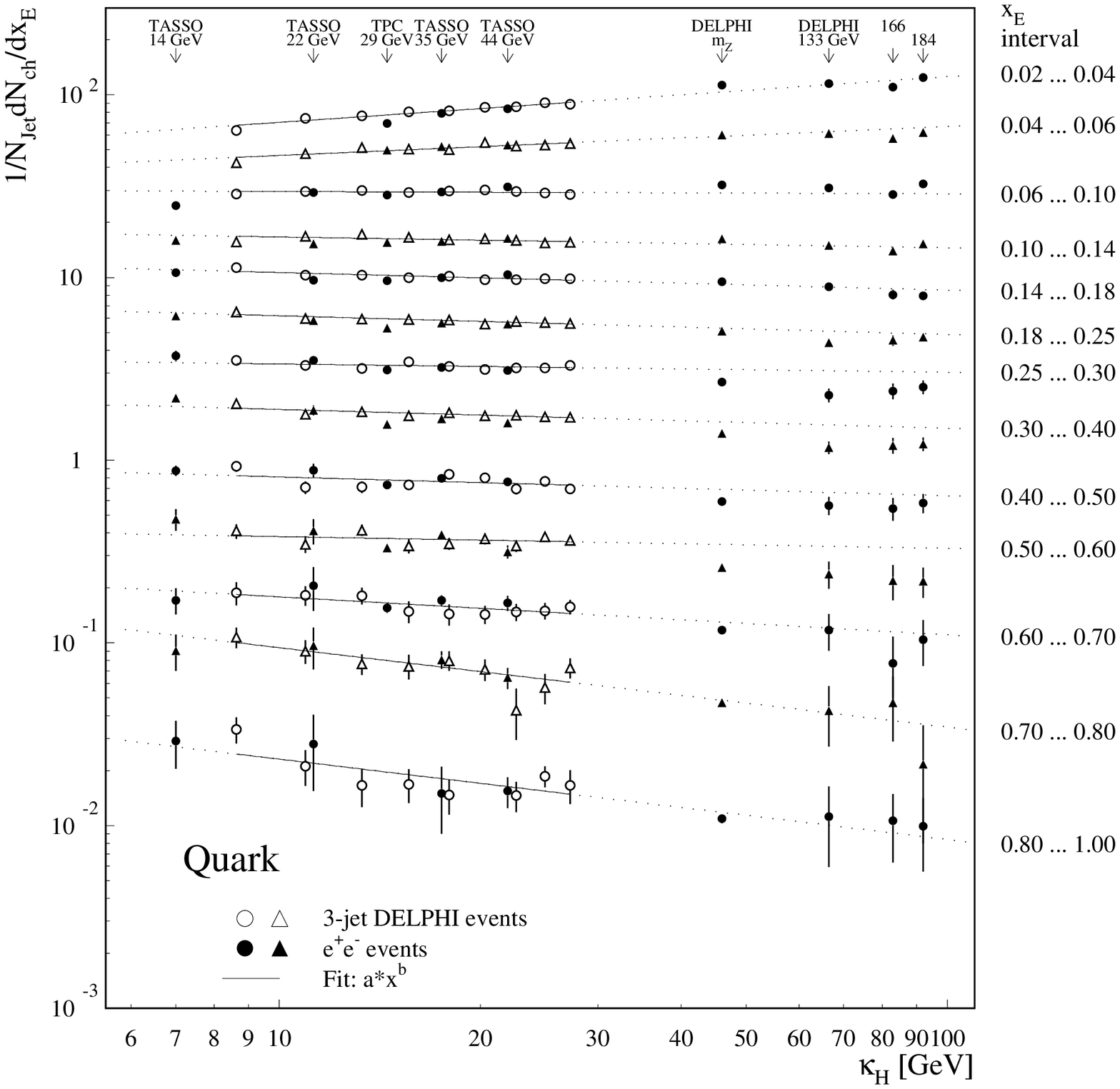,%
width=6.8cm,height=7.5cm,bbllx=0pt,bblly=0pt,bburx=565pt,bbury=565pt,clip=}
\epsfig{file=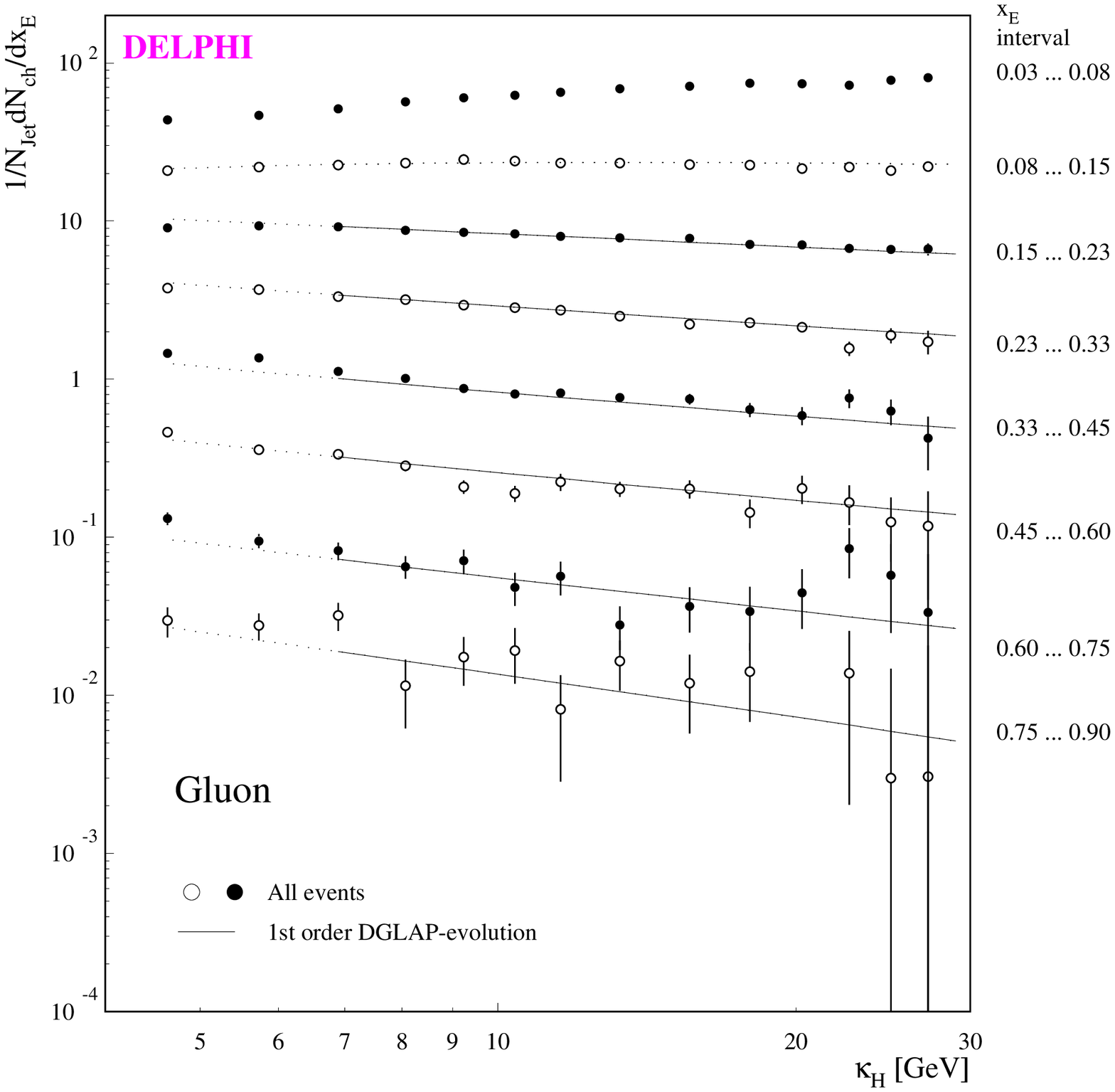,%
width=6.8cm,height=7.5cm,bbllx=0pt,bblly=0pt,bburx=570pt,bbury=570pt,clip=}
\vspace*{-0.2cm}
\caption{\small Scale dependence of quark and gluon jet fragmentation 
function. Left: The data from lower energy experiments are multiplied by 0.5, 
since these refer to the multiplicities in $q\bar{q}$ events rather than in a
single quark jet; $Q=E_{c.m.}/2$.}
\label{ffscal}
\end{figure}
In Fig.\ref{ffincl} the inclusive quark fragmentation function is compared 
to the gluon one. The latter is observed to be softer which can be explained 
by the fact that the radiation of soft gluons is larger for gluon jets and 
that gluon cannot be present as a valence parton inside a produced hadron 
(first splitting $g\rightarrow q\bar{q}$ has to occur). 
The scale dependence of the quark and gluon fragmentation functions is 
presented in Fig.\ref{ffscal}. The figure on the left contains a summary 
of quark jet
fragmentation function measurements. A good correspondence between the biased 
and unbiased measurements suggests that $\kappa_H$ is a meaningful choice of
scale for a general 3-jet topology. The figure on the right shows the
scaling violations of the biased gluon jets which are stronger than
for quark jets. This is due to the fact that the scale dependence of the 
fragmentation functions for gluons is dominated by the splitting 
$P_{g\rightarrow gg} \sim C_A$, while that for quarks is dominated by the
splitting $P_{q\rightarrow qg} \sim C_F$.
\vspace*{-0.5cm}
\section{Conclusions}\vspace*{-0.2cm}
Shown examples of differences between quark and gluon jets underline the key
role of the inequality ${\mathbf C_A>C_F}$. Its consequences, namely larger 
widths and multiplicities as well as softer fragmentation function with 
stronger scaling violations of gluon jets with respect to quark jets have been
confirmed experimentally. 

A new method for the indirect measurement of unbiased $N_{gg}$ from biased 
$N_{q\bar{q}g}$ and $N_{q\bar{q}}$ was described and its usefulness proven. 
The results for $N_{gg}$ based on $k_{\bot,Lu}$ agree significantly better 
with previous measurements and MC predictions than the results based on 
$k_{\bot,Le}$. An overall conclusion is that the theory is in general
agreement with the experimental results.
\vspace*{-0.4cm}  
\subsection*{Acknowledgements}\vspace*{-0.2cm}
I wish to thank William Gary and Joost Vossebeld for their help.

\end{document}